# A Short Review of Plasmonic Graphene-Based Resonators: Recent Advances and Prospects

**Mohammad Bagher Heydari** [1,*], **Mohammad Hashem Vadjed Samiei** [1]

[1] School of Electrical Engineering, Iran University of Science and Technology (IUST), Tehran, Iran

[*]Corresponding author: mo_heydari@elec.iust.ac.ir

**Abstract:** This article aims to study graphene-based resonators published in the literature. Graphene resonators are designed based on graphene conductivity, a variable parameter that can be changed by either electrostatic or magnetostatic gating. A historical review of plasmonic graphene resonators is presented in this paper, which can give physical insight to the researchers to be familiarized with four types of graphene-based resonators: 1-Ring resonators, 2- planar, 3- Fabry-Perot, and 4-other resonators which are not categorized in any of the other three groups.

## 1. Introduction

In recent years, plasmonics is a new emerging science, which exhibits novel features in the THz region [1, 2]. Both metals and two-dimensional (2D) materials can support surface plasmon polaritons (SPPs). In metals, SPPs often are excited in the visible and near-infrared regions [3, 4]. Graphene is a 2D sheet that offers fascinating properties. One of these properties is optical conductivity, which can be tuned by either electrostatic or magnetostatic gating or via chemical doping. This fascinating feature makes graphene a good candidate for designing innovative THz components such as waveguides [5-11], filters [12], couplers [13], Radar Cross-Section (RCS) reduction-based devices [14-16], and graphene-based medical components [17-23]. Moreover, the hybridization of graphene with anisotropic materials such as ferrites, which have many features [24, 25], can enhance the performance of the graphene-based devices [6].

In this paper, we study plasmonic graphene-based resonators. In section 2, various types of graphene-based resonators are reviewed in detail. These resonators have four categories: ring resonators, planar resonators, Fabry-Perot resonators, and other resonators. In all of these resonators, the graphene has been biased electrically. Finally, section 3 concludes the paper.

## 2. Various Types of Plasmonic Graphene-Based Resonators

In this section, we consider graphene-based resonators in four groups: 1-Ring resonators, 2- planar, 3- Fabry-Perot, and 4-other resonators which are not categorized in any of the other three groups.

### 2.1 Ring Resonators

A ring resonator is a kind of waveguide in which the input power is coupled to at least one closed-loop and get out of the output bus. In the optics region, the optical ring resonators work based on one of the three main principles: 1- optical coupling, 2- total internal reflection, and 3- interference. Ring resonators based on graphene have been studied in numerous articles since 2013. Lei Zhang et al have proposed a compact ring resonator that works based on edge SPP modes of graphene disk [26]. The resonator is composed of a graphene disk with a radius of *R* and a graphene ribbon with a width of *w*, as seen in fig. 1 [26]. Incident power from the graphene ribbon coupled to graphene disk when momentum match between SPP waves of ribbon and disk occurs [26]. The structure is simulated in COMSOL software and the transmission spectrum indicates a series of dips which means that the structure operates as a resonator [26].

Rai Kou and his coworkers have fabricated a silicon ring resonator integrated with graphene [27, 28]. Fig. 2 illustrates the top view and cross-view of the proposed ring resonator. The authors focus on quality factor (Q-factor) variations of the resonator in presence of graphene [27]. It has been reported that by changing the graphene length from 0 to 20 $\mu m$, the Q-factor decreases from 7900 to 1200 [27].

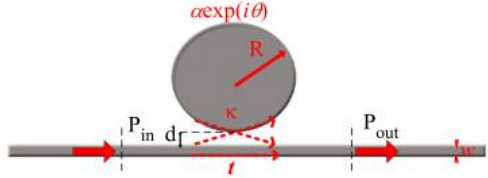

Fig. 1. The plasmonic resonator composed of a GNR waveguide and graphene disk [26]

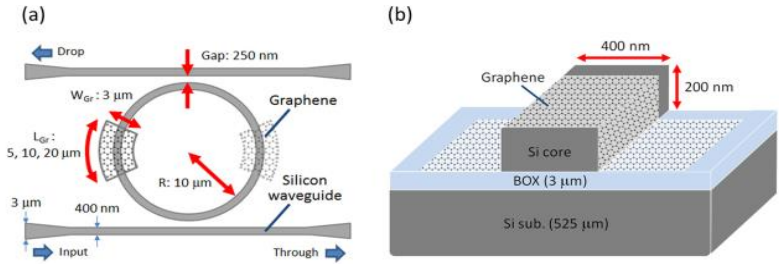

Fig. 2. (a) Top view, (b) cross-sectional view of silicon-graphene ring resonator [27]

The sharp transparency window which has a broad absorption spectrum in the medium is called electromagnetically induced transparency (EIT). This phenomenon occurs when atomic upper levels and applied optical field have coherent interactions with each other. Recently, an analogy phenomenon to EIT, known as plasmonically induced transparency (PIT) has appeared. Jicheng Wang and his coworkers have been studied PIT in graphene ring resonators in 2015 [29-33]. All structures introduced by this research group are analyzed by coupled-mode theory (CMT) [29-33]. In [29, 30], they investigate the PIT in a plasmonic resonator composed of graphene waveguide and two graphene rings on one side, as exhibited in fig. 3 (a) [29, 30]. The graphene rings and graphene waveguide have different bias voltage from each other [29, 30]. At first, the authors have described near field coupling systems for small width of $w$ [29, 30]. In this case, phase coupling cannot affect resonance frequency [29, 30]. The transmission spectrum for near field coupling indicates that as $w$ decreases from 80 nm to 20 nm, the PIT windows become more apparent [29, 30]. Then, the study of phase coupling systems for large separation $w$ has been considered in the rest of the paper [29, 30]. Unsymmetrical graphene ring resonators were introduced in [33]. Fig. 4 (b),(c) display two unsymmetrical ring resonators which are analyzed by CMT in the article [33]. By using three rings resonator, multiple new PIT windows can be appeared [33].

Jicheng Wang's research group has designed two applied structures based on ring resonators [30, 31]. The first applied structure is Mach-Zehnder Interferometer (MZ-I) constructed by two graphene sheets as input and output and two graphene rings which are located between the input and output graphene sheets [30]. Indeed, this MZ-I is a bandpass filter which its passing frequency can be changed by varying the chemical potential of graphene rings [30]. The second useful structure is Multi-channel Demultiplexer, as demonstrated in fig. 4(a) [31]. Fig. 4 (b)-(d) show the normalized electric field distribution at $7.28\,\mu m$, $7.66\,\mu m$ and $6.94\,\mu m$, respectively. It is obvious that each ring resonator in the demultiplexer structure works at a special resonance frequency and then the input power can couple to each channel at a specified wavelength [31].

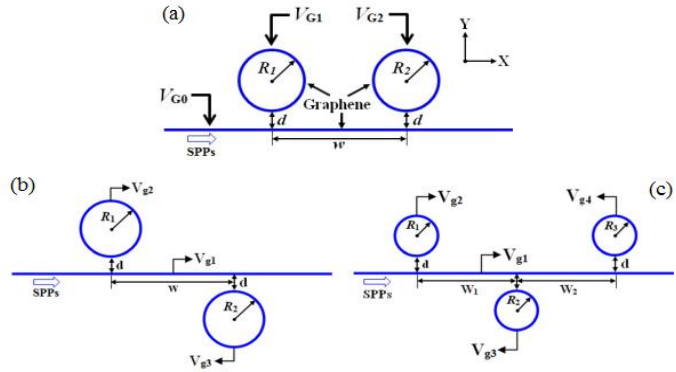

Fig. 3. Multi-mode PIT resonator consists of graphene waveguide and: (a) two graphene rings [29, 30], (b) two graphene rings at both sides [33], (c) three graphene rings [33]

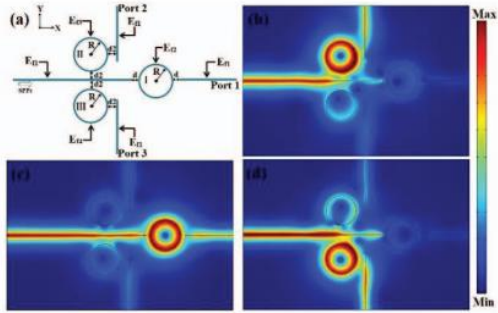

Fig. 4. (a) The Multi-channel demultiplexer based on ring resonators, (b)-(d) The normalized electric field distribution at $7.28\,\mu m, 7.66\,\mu m$ and $6.94\,\mu m$, respectively [31]

C.Ciminelli et al are designed a novel graphene-based resonator in [34]. The graphene is utilized to tune the properties of the resonator cavity which can be utilized as a phase shifter, delay line, or a switch [34]. An accurate study on graphene split-ring resonators has been done in [35]. In this research, the authors utilized three graphene split-ring resonators and a nano-ribbon waveguide to design an electrically controllable switch in the THz regime [35]. The schematic of the mentioned structure is displayed in fig. 5 [35]. The authors applied a unitary scattering matrix to derive an analytical model for the structure [35]. The effect of varying $\mu_c$ on extinction ratio has been discussed in the context of this paper and the output results indicate that the extinction ratio of the switch can be tuned as 30-0.82 dB by changing chemical potential from 0.1-0.2 eV [35].

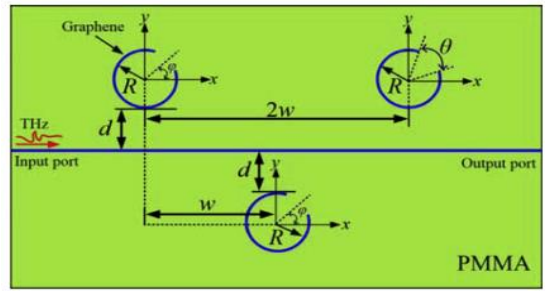

Fig. 5. The schematic of plasmonic switch or attenuator composed of three graphene split rings and graphene waveguide [35]

A novel graphene ring resonator is presented in [36]. The resonator is composed of a hybrid plasmonic bus waveguide and a ring sensor where the resonator has a $Si_3N_4$ layer as a dielectric layer and a graphene strip separated by a $SiO_2$ layer [36]. The cladding is porous- alumina (p-$Al_2O_3$) and the plasmonic waveguide has a similar structure to the ring resonator [36]. By using FDTD, the structure numerically analyzed and results indicate that the hybrid ring resonator has high FOM in comparison to a similar optical structure [36]. Thomas Christopoulos et al. developed CMT for graphene nonlinear resonators in [37]. In this article, the authors focused on third-order nonlinearity and presented an efficient framework for modeling nonlinear material such as graphene [37]. To validate the analytical model, a graphene-based resonator was chosen where its structure was similar to fig. 1 [37].

Several papers about the graphene ring resonators have been published in 2017. Yaun Meng et al. have introduced a compact graphene ring resonator works as an optical router [38], which its structure is similar to the studied structure in [27, 28]. The designed router has unique features such as tunability with varying chemical potential, low cross-talk (-15 dB at a wavelength interval of 5 nm), and low energy consumption (59 fJ/bit) which makes it a promising candidate for future photonic devices [38].

A novel crescent resonator (CR) based on graphene has been reported in [39]. The CR has much-improved field enhancement than a conventional split-ring resonator [39]. Fig. 6 represents the structure of hybrid metal-graphene CR where is composed of metal crescent resonator deposited on graphene-dielectric-metal layers [39]. This hybridization helps to dynamically control SPP tuning and high field enhancement is achieved [39]. The authors investigate the field distribution in hybrid resonators for various $\mu_c$ to enhance the field concentration in the structure [39]. The role of the graphene sheet in field enhancement and tunability of the CR has been discussed more precisely in the article [39].

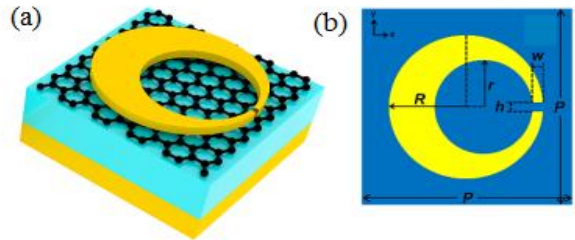

Fig. 6. (a) 3D view, (b) cross-section view of hybrid CR based on hybridization of metal-graphene [39]

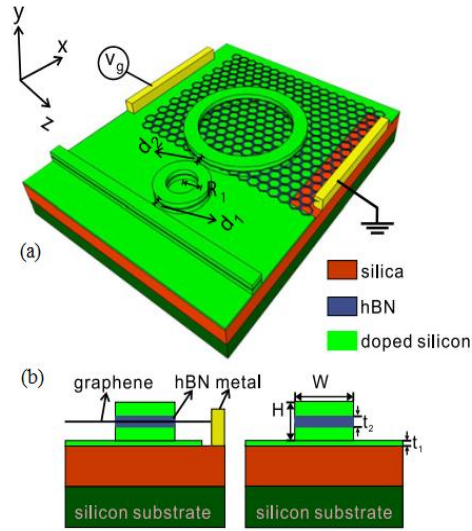

Fig. 7 (a) 3D view of two coupled ring resonators,
(b) Cross sections of ring resonator (left panel) and bus waveguide (right panel) [40]

Two coupled ring resonators, which one of them is placed on the graphene sheet have been studied by Xuetong Zhou et al. to achieve tunable electromagnetically induced transparency (EIT), as exhibited in fig. 7 [40]. It should be noted that the proposed device is tunable by varying the chemical potential of graphene or gate voltage which means that EIT resonance and effective refractive index can be tuned to obtain variable output power [40]. Patrizia Savi and his coworkers have fabricated a new graphene slotted ring resonator which can be utilized for sensing [41]. This paper minimizes the RF losses by capacitively loading the ring at selective locations [41].

*2.2 Planar Resonators*

Planar graphene-based resonators will be reviewed in the literature in this section. Xu Han et al. report novel planar GNR coupled with graphene rectangular resonators for tunable PIT [42]. Fig. 8 illustrates the proposed resonators where only TM SPP waves are studied in this work [42]. The width of graphene nano-ribbon waveguide is assumed *w=10 nm* and all of the structures are located on a sapphire substrate with a thickness of *50 nm* [42]. All structures are simulated in Numerical FDTD solutions software. The structure of fig.8 (d) is proposed to obtain a double-channel PIT effect [42]. Furthermore, two techniques are applied to achieve the PIT effect which is the near-field coupling between the radiative and dark resonator and the indirect coupling through the GNR waveguide [42]. In the article, the transmission spectrum for various $\mu_c$ and distributions of magnetic and electric fields for all structures are depicted and have been discussed more precisely [42]. The more explanations exist in the context and the reader can refer to the article [42].

Plasmonic anti-symmetric coupling resonator based on parallel interlaced graphene pair is suggested by Hong-Ju Li et al. and studied numerically by FDTD [43]. These resonators consist of parallel double-layer graphene in the air and two spatial silver strips with a width of *w*, as seen in fig.9 [43]. The distance between two silver strips is assumed *L*. It should be noted that silver has complex relative permittivity described by Drude's model [43]. FDTD results demonstrate two transmission peaks were appeared in 8.58 and 7.15 $\mu m$, as demonstrated in fig. 10 (a) [43]. For wavelength 8.58 $\mu m$, contour profiles of $H_z$ are depicted in fig. 10 (c) [43]. It is obvious that the most SPP power is coupled to the resonator and transmits the right graphene sheet [43]. As represented in fig. 10 (b), the incident wave cannot be coupled and reflected because it does not satisfy the resonance condition [43]. In the rest of the paper, the authors have focused on parametric study. For instance, they have investigated the influence of increasing the distance between two silver strips (L) on the transmission spectrum [43].

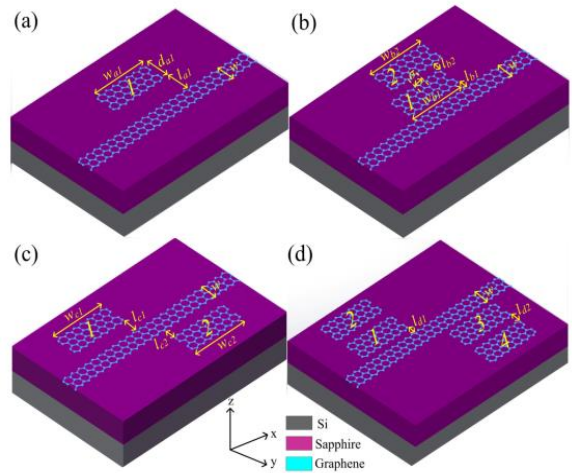

Fig. 8. (a) One resonator coupled structure, (b,c) two-resonator coupled structure, (d) four-resonator coupled system[42]

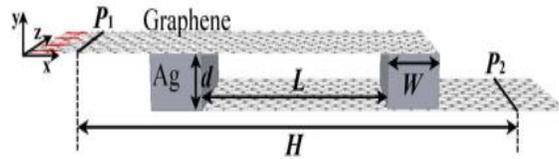

Fig. 9. The schematic of anti-symmetric coupling resonator based on parallel interlaced graphene pair [43]

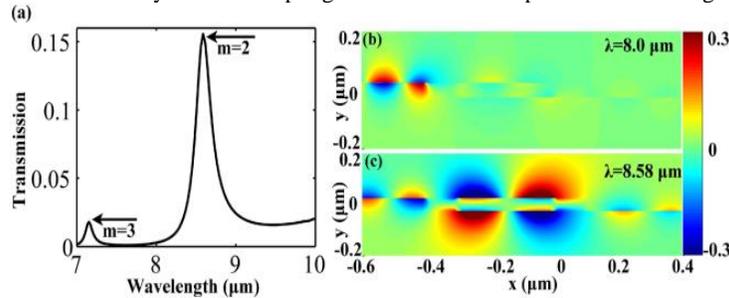

Fig. 10. (a) The transmission spectrum of the proposed anti-symmetric coupling resonator as a function of wavelength, $H_z$ field distributions at: (b) $\lambda = 8 \mu m$, (c) $\lambda = 8.58 \mu m$ [43]

Novel THz resonator based on multi-layer graphene ribbons is introduced by Xiang Li in [44]. The authors suggest an equivalent circuit model (ESC) for their resonator that can be determined the resonance frequency and unloaded Q-factors [44]. The mathematical relations of this model have been derived and given in [44]. Simulated results exhibit that performance of the resonators is much better than the copper and neutral graphene band counterparts. For instance, a high Q-factor of 80.6 at *f=793.1 GHz* has been reported for this resonator [44].

Jicheng Wang and his coworkers have designed graphene sheet resonators to achieve active multiple PIT in [45]. The proposed structure is composed of the GNR as the main waveguide, two graphene sheets as resonators where are located at a distance of *t* from the main waveguide, and a distance of *l* from each other [45]. The authors consider the proposed structure as a plasmonic filter and then the phase coupling between two graphene sheet resonances is studied in the paper [45].

Nan Chen et al. have been presented a graphene-based plasmonic resonator at the 2016 international conference on optics [46]. In the mentioned paper, they experimentally studied the plasmonic resonator and emphasized that their proposed resonator can be applied in sensing applications [46]. Hadiseh Nasari et al. focus on nonlinear optics of SPPs for designing graphene ribbon resonators [47]. The studying structure has two semi-infinite graphene sheets as a waveguide and a graphene ribbon of length L as a resonator and the whole structure is placed on a $SiO_2/Si$ substrate, as shown in fig. 11 [47]. The authors have characterized the nonlinear response of their suggested structure by optical Kerr effect in the substrate and numerical results have been illustrated in the article [47].

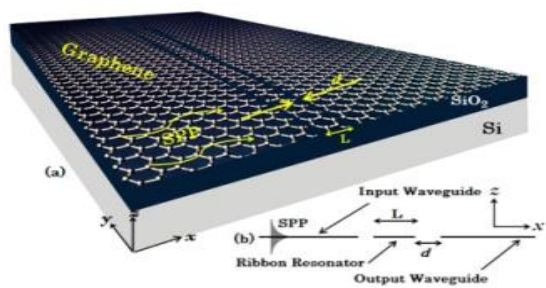

Fig. 11. (a) 3D view, (b) Side view of planar studied structure [47]

*2.3 Fabry-Perot Resonators*

In this section, a short review of graphene Fabry-Perot Resonators (FPR) will be presented. As exhibited in fig.12 (a), the proposed FPR is composed of two metallic mirrors and graphene has been located between two cavities [48]. The bottom mirror is supposed gold as a back-gate for graphene, while the top Bragg mirror is composed of a period low ($n_L$) and high refractive index dielectric ($n_H$) where *N* is the number of bilayers [48]. The thicknesses of these dielectrics satisfy $n_H t_H = n_L t_L = \lambda_0/4$ which $\lambda_0$ is the wavelength of the lowest resonator cavity mode [48]. The author reported the insertion loss about zero and the modulation depth of 100% for impedance matching of resonators [48].

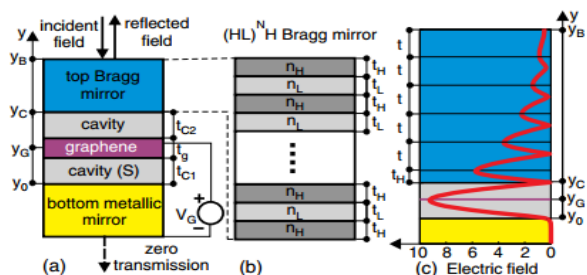

Fig. 12. (a) The schematic of FPR, (b) The schematic of top Bragg mirror, (c)Typical electric field distribution [48]

Huawei Zhuang and his coworkers have been investigated graphene-based EIT by using FPR and introduced applied devices based on FPR [49, 50]. Fig. 13. (a) exhibits the GNR waveguide coupled to FPR where the gap between GNR waveguide and FPR is *t* [49]. The transmission spectrum for geometrical parameters of $w=20 nm, L=90 nm, t=5 nm$ and distribution of $E_z$ at *f=11.279 THz* are displayed in fig. 13 (b) [49]. In the rest of this paper, the authors study GNR waveguide coupled to two FPRs more precisely and perform a parametric study to consider the influence of various parameters on the transmission spectrum [49]. Huawei Zhang et al. introduce a novel wavelength demultiplexer (WDM) based on FPR by the FDTD method in [50]. The authors used a temporal CMT to describe the physical characteristics and aspects of the WDM structure [50]. A detailed explanation is reported in [50].

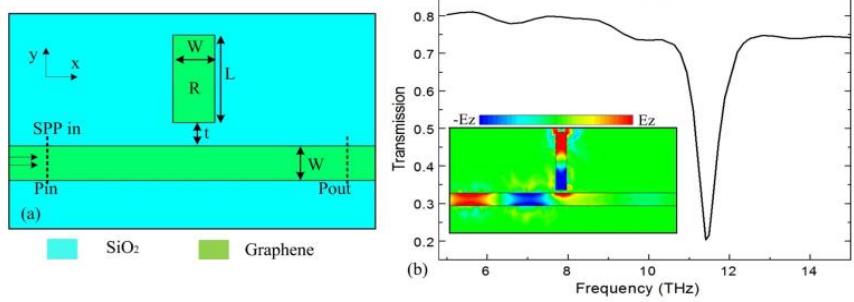

Fig. 13. (a) The schematic of GNR coupled to one FPR, (b) Transmission spectrum as a function of wavelength and $E_z$ field distribution at $f=11.279$ $THz$ for geometrical parameters of $w=20\,nm, L=90\,nm, t=5\,nm$ [49]

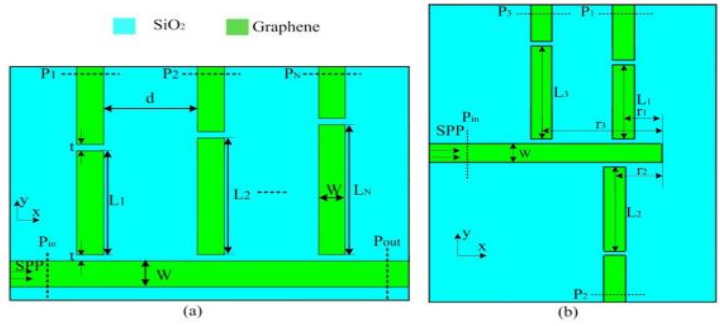

Fig. 14. (a) The schematic of $1\times N$ WDM based on FPR, (b) The schematic of triple-WDM based on FPR [50]

*2.4 Other Resonators*

This section studies two various graphene-based resonators that are not categorized in any of the mentioned groups. The article [51] investigates graphene hybrid plasmonic resonator used in nano-laser. The proposed resonator is illustrated in fig. 15 (a) where a metal disk is located on top of the dielectric slab and the graphene layer has been sandwiched between InGaAs and SiO$_2$ dielectric layers [51]. Graphene has been utilized in this structure to enable modifying the resonant mode spectrum [51]. The hybrid SPP forms whispering-gallery modes due to the cylindrical symmetry of the resonator [51]. The resonator has been simulated by FDTD and results indicate that the dominant mode can be varied in a wide range of *750 nm* wavelength by changing Fermi level from *0.48-85 eV* [51].

In [52], a graphene-based waveguide resonator for submillimeter-wave applications at 100-1100 GHz has been designed and studied. The resonator is composed of two symmetrically placed E-plane inserts where a part of the inserts is covered by graphene [52]. The authors first analyzed the structure and then consider the dependence of the resonator properties such as Q-factor to various parameters [52]. This device can be a good candidate for the compact components in the sub-millimeter region due to its tunability [52].

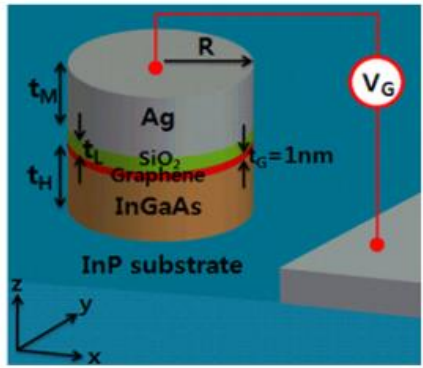

Fig. 15. The schematic of hybrid metal-graphene cylindrical resonator [51]

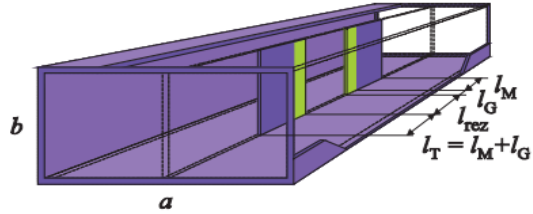

Fig. 16. The Graphene-based resonator waveguide consists of two equal placed *E*-plane inserts [52]

## 3. Conclusion

We introduced and studied plasmonic graphene-based resonators published in the literature in this article. The graphene-based resonators are divided into four main categories: 1- Ring resonators, 2- planar, 3- Fabry-Perot, and 4- other resonators. In the fourth category, we reviewed all resonators which are not categorized in any of the three groups. In all of these resonators except some structures (non-linear graphene resonators) mentioned in the context, the graphene was biased electrically.